\documentclass[twocolumn]{aastex631}

\usepackage{graphicx}
\usepackage{amsmath,amsfonts,amsthm}
\usepackage{aas_macros}
\usepackage{nicefrac}

\usepackage{savesym}
\savesymbol{tablenum}

\usepackage{siunitx}
\DeclareSIUnit\year{yr}
\DeclareSIUnit\parsec{pc}
\DeclareSIUnit\gauss{G}

\begin{document}
\reportnum{TTK-24-33}

\title{Investigating the CREDIT history of supernova remnants as cosmic-ray sources}

\author[0000-0001-6618-9684]{Anton Stall}
\correspondingauthor{Anton Stall}
\email{stall@physik.rwth-aachen.de}

\author[0009-0006-0880-4865]{Chun Khai Loo}
\email{khai.loo@rwth-aachen.de}

\author[0000-0002-2197-3421]{Philipp Mertsch}
\email{pmertsch@physik.rwth-aachen.de}

\affiliation{Institute for Theoretical Particle Physics and Cosmology (TTK), RWTH Aachen University, 52056 Aachen, Germany}

\begin{abstract}
Supernova remnants (SNRs) have long been suspected to be the primary sources of Galactic cosmic rays. 
Over the past decades, great strides have been made in the modeling of particle acceleration, magnetic field amplification, and escape from SNRs. 
Yet while many SNRs have been observed in nonthermal emission in radio, X-rays, and gamma rays, there is no evidence for any individual object contributing to the locally observed flux. 
Here, we propose a particular spectral signature from individual remnants that is due to the energy-dependent escape from SNRs. 
For young and nearby sources, we predict fluxes enhanced by tens of percent in narrow rigidity intervals; given the percent-level flux uncertainties of contemporary cosmic-ray data, such features should be readily detectable.  
We model the spatial and temporal distribution of sources and the resulting distribution of fluxes with a Monte Carlo approach. 
The decision tree that we have trained on simulated data is able to discriminate with very high significance between the null hypothesis of a smooth distribution of sources and the scenario with a stochastic distribution of individual sources. 
We suggest that this cosmic-ray energy-dependent injection time (CREDIT) scenario be considered in experimental searches to identify individual SNRs as cosmic-ray sources.  
\end{abstract}

%----------------------------------------------------------
%----------------------------------------------------------
%----------------------------------------------------------
\section{Introduction}

Cosmic rays with energies up to at least $E_{\text{knee}} \simeq \SI{3}{\peta\electronvolt}$ are usually assumed to be of Galactic origin~\citep{ParticleDataGroup:2024cfk,Gabici:2019jvz,Kachelriess:2019oqu}. 
Supernova remnants (SNRs)~\citep{2008ARA&A..46...89R} have long been considered the prime candidate for the origin of these Galactic cosmic rays (GCRs) for a number of reasons: 
first, observations of SNRs oftentimes show power-law spectra, e.g., in radio~\citep{2019JApA...40...36G}, X-rays~\citep{ChandraSupernovaCatalogue} or gamma rays~\citep{2016ApJS..224....8A}. 
Second, diffusive shock acceleration (see, e.g.,~\citet{2001RPPh...64..429M}) can operate at the SNRs' blast waves, thus providing a likely acceleration mechanism. 
Finally, the locally observed energy density of cosmic rays can be explained if on average $\approx 10\%$ of the shock kinetic energy is converted to GCRs~\citep{1964ocr..book.....G}. 
Yet observational evidence for SNRs to accelerate cosmic rays up to $E_{\text{knee}}$ is scarce~\citep{Gabici:2019jvz}.

From a modeling perspective, it is not clear how SNRs can accelerate particles to a few petaelectronvolts either. 
Adopting shock speeds of $\mathcal{O}(10^4) \, \si{\kilo\metre\per\second}$ and a (turbulent) magnetic field strength typical of the interstellar medium, i.e., $\approx \SI{1}{\micro\gauss}$, it can be shown~\citep{1983LagageCesarskyA&A} that the maximum particle energy is of the order of \SI{10}{\tera\electronvolt}, which falls short of $E_{\text{knee}}$ by some three orders of magnitude. 
Attaining energies of $E_{\text{knee}}$ instead would require significant amplification of the turbulent magnetic field. 
This amplification is believed to be achieved thanks to the non-resonant hybrid instability (also known as Bell instability)~\citep{2004BellMNRAS}, which is driven by the macroscopic current of escaping cosmic rays. 

In the following, we lay out the concordance view of time-dependent acceleration at and escape from SNR shocks, e.g.,~\citep{2009CaprioliBlasiMNRAS2,Schure:2013kya,Bell:2013kq,Cristofari:2020mdf, 2009GabiciAharonianMNRAS,2012ThoudamHorandelMNRAS,2019CelliMorlinoMNRAS}. 
The amplification of magnetic fields is time-dependent due to its dependence on environmental parameters, most importantly the shock speed.
The non-resonant hybrid instability operates most efficiently at high shock speeds, which can be achieved in the ejecta-dominated phase of the SNR evolution.
The maximum energy, or better rigidity $\mathcal{R}_{\text{max}}$,~\footnote{Rigidity is defined as $\mathcal{R} = {p c}/{Z e}$ where $p$ denotes the particle's momentum, $c$ the speed of light, $Z$ the charge number, and $e$ the unit charge.} is expected to be attained before $t_{\text{Sed}}$, the start of the Sedov-Taylor phase, when the amount of matter accumulated by the shock equals the ejecta mass.
For $t > t_{\text{Sed}}$, the shock speed decreases and so does the maximum rigidity $\mathcal{R}_{\text{max}}(t)$.
Only particles of rigidities $\mathcal{R} < \mathcal{R}_{\text{max}}(t)$ can be confined by the magnetized turbulence across the shock, and so particles of rigidity $\mathcal{R}_{\text{max}}(t)$ escape at time $t$, starting with the highest-rigidity particles at $t_{\text{Sed}}$. 
The trend continues down to a rigidity $\mathcal{R}_{\text{b}}$ (e.g., \SI{10}{\tera\volt}), below which particles can be confined in the vicinity of the shock throughout the Sedov-Taylor phase without magnetic field amplification.
Hence, all particles with $\mathcal{R} < \mathcal{R}_{\text{b}}$ escape at approximately the same time $t_{\text{life}}$ when the SNR shock dissipates.
One could parameterize this time-dependence of cosmic-ray escape with a source term, that is, the number of GCRs released from a source per unit time and rigidity, of the form $Q(\mathcal{R}, t) \equiv \delta(t - t_{\text{esc}}(\mathcal{R})) Q_{\mathcal{R}}(\mathcal{R})$ where $t_{\text{esc}}(\mathcal{R})$ is the inverse function of $\mathcal{R}_{\text{max}}(t)$ and $Q_{\mathcal{R}}(\mathcal{R})$ is an arbitrary function of rigidity but likely close to power-law form. 
Similar concordance scenarios have been widely adopted in the literature and are also supported by numerical simulations~\citep{2012BlasiAmatoJCAP,2012ThoudamHorandelMNRAS,2009CaprioliBlasiMNRAS1}. They have been used to study gamma-ray emissions in the vicinity of cosmic-ray sources as well~\citep{2009GabiciAharonianMNRAS,2019CelliMorlinoMNRAS}. 
We will refer to our concordance model of a time-dependent injection of cosmic-ray particles above $\mathcal{R}_{\text{b}}$ as the \textit{Cosmic-Ray Energy-Dependent Injection Time} (CREDIT) scenario. 

The details of the cosmic-ray escape are usually not considered when trying to explain the locally observed GCR measurements. 
Instead, the rate of injection of GCRs into the interstellar medium is modeled as the product of a smooth source density $n_{\text{src}}(\vec{r})$ and a steady rate per unit rigidity $\bar{Q}(\mathcal{R})$. 
This rate can be considered the temporal average of the time-dependent source term $Q(\mathcal{R}, t)$~\citep{Cristofari:2020mdf}:
\begin{equation}\label{eq:time_inegrated_source_spectrum}
\bar{Q}(\mathcal{R}) \equiv \nu \int_0^{\infty} \mathrm{d} t \, \delta(t - t_{\text{esc}}(\mathcal{R})) Q_{\mathcal{R}}(\mathcal{R}) = \nu Q_{\mathcal{R}}(\mathcal{R}) \, .
\end{equation}
Here, $\nu$ is the Galactic rate of supernovae. 

It is hence natural to ask if there are observable consequences of the CREDIT scenario for locally measured cosmic rays. 
If one adopts a distribution of sources that is spatially smooth and steady, as is usually assumed, e.g., in numerical finite difference codes (see, however, \citet{Porter:2019wih}) then there are none, that is, observationally the time-dependent escape scenario cannot be distinguished from a time-independent scenario. 

In reality, SNRs are individual objects with spatial and temporal extents much smaller than the distances and times over which cosmic rays propagate. 
The smooth and steady assumption can be justified in situations where a large number of sources contribute to the flux of GCRs.
In such a situation, the measured flux for a particular distribution of sources will differ little from the mean of an ensemble of such source distributions. 
This mean is identical to the prediction from the smooth source distribution due to the linearity of GCR transport.
For instance, it is easy to estimate that for standard parameters of GCRs, the number of sources contributing to the flux at \SI{10}{\giga\volt} is about 100 times larger than at \SI{100}{\tera\volt}, where only some tens of sources cause the bulk of the flux. This justifies the use of the smooth scenario at gigaelectronvolt energies. 

There are other situations, however, when the assumption is not justified. 
Examples include the transport of GCR electrons and positrons above a few \SI{100}{\giga\electronvolt}~\citep{Pohl:1998ug,2018MertschJCAP} or of GCRs below \SI{1}{\giga\volt}~\citep{2021PhRvL.127n1101P}; in both cases, the ranges are limited by energy losses. 
Consequently, the spectrum becomes very sensitive to the actual distribution of sources. 
For certain realizations of the source distribution, this can lead to deviations from the prediction of a smooth model and might be observable. 
Given our limited knowledge of the true source distances and ages, we have to resort to Monte Carlo simulations in which the total flux is simulated for different realizations of sources drawn from a smooth distribution~\citep{Mertsch2011,Genolini2017,EvoliAmatoBlasi2021_Stochastic,2012BlasiAmatoJCAP}. 
We refer to models of this sort as stochastic models. 

In the following, we argue that the CREDIT scenario put into the context of a stochastic model can give large and, more importantly, observable spectral features at teravolt rigidities. 
We illustrate this by computing the fluxes of GCR protons for a large ensemble of source distributions in the framework of a realistic, yet simple cosmic-ray model. 
We characterize the spectral features and employ machine learning to reliably discriminate the spectral fluctuations in a CREDIT scenario from the fluctuations due to statistical errors in a smooth model.
We find that identifying signatures of discrete source modeling should be within reach of current experiments given their unprecedented accuracy. This makes it possible to directly test the supernova paradigm.

%----------------------------------------------------------
%----------------------------------------------------------
%----------------------------------------------------------
\section{Methodology}

To study the influence of source stochasticity on the locally observed GCR spectrum, we consider a simplified model of the Galaxy. We assume that all sources lie in the Galactic disk ($z=0$) and cosmic rays diffuse through the Galactic halo that extends to $z=\pm H$.
We draw the source ages from a uniform distribution adopting the canonical supernova rate $\nu$ of \SI{0.03}{\per\year}~\citep{BerghTammann1991_SNRrate,Tammann1994_SNRrate} and the source positions from an axisymmetric distribution just depending on galactocentric distance~\citep{Ferriere2001}, assuming that the Sun is located at radius $R_{\odot}$.

We will only consider GCR protons with rigidities above some gigavolts such that inelastic collision and advection can be ignored. We also assume an isotropic diffusion coefficient that follows a power law in rigidity~\footnote{$\beta$ is the ratio of the particle's speed to the speed of light.} $\kappa\left(\mathcal{R}\right) = \kappa_0 \beta\left(\mathcal{R}\right) \left(\mathcal{R}/\mathcal{R}_0\right)^{\delta}$. So, the transport equation is
\begin{equation}\label{eq:transport_equation}
    \frac{\partial \psi_{\mathcal{R}}\left(\mathcal{R}, t, \mathbf{x}\right)}{\partial t} - \kappa\left(\mathcal{R}\right) \nabla^2 \psi_{\mathcal{R}} \left(\mathcal{R}, t, \mathbf{x}\right) = Q\left(\mathcal{R}, t, \mathbf{x}\right) \, ,
\end{equation}
where $\psi_{\mathcal{R}}\left(\mathcal{R}, t, \mathbf{x}\right)= {\mathrm{d} n}/{\mathrm{d} \mathcal{R}}$ denotes the isotropic CR density ($n$ is the number density). It is related to the differential flux $\Phi_{\mathcal{R}} = (\mathrm{d}^4n)/(\mathrm{d}\mathcal{R} \, \mathrm{d}A \, \mathrm{d}t \, \mathrm{d}\Omega) = v/(4 \pi) \ \psi_{\mathcal{R}}$ and to the phase-space density $f=f\left(p, t, \mathbf{x}\right) $ through $\psi_{\mathcal{R}} = (4 \pi p^2 Z e)/c \ f$.
 
We assume that a single population of SNRs injects all GCR protons up to the knee rigidity $\mathcal{R}_{\text{knee}}$ 
to keep the complexity of this study low.
In the CREDIT scenario, the particles' escape times are rigidity-dependent. This escape time $t_{\text{esc}}(\mathcal{R}) = t_0 + \Delta t_{\text{esc}}(\mathcal{R}) $ depends on the time of the supernova explosion $t_0$ and the time it takes a proton of a certain rigidity to escape the source $\Delta t_{\text{esc}}(\mathcal{R})$. We assume that the latter follows a power law at high rigidities (as indicated by~\citet{2009CaprioliBlasiMNRAS1}):
\begin{equation}
    \Delta t_{\text{esc}}\left(\mathcal{R}\right) = t_{\text{Sed}} \!\left(\frac{\mathcal{R}}{\mathcal{R}_{\text{knee}}}\right)^a \, , \; a = \frac{\ln\left(t_{\text{life}}/t_{\text{Sed}}\right)}{\ln\left(\mathcal{R}_{\text{b}}/\mathcal{R}_{\text{knee}}\right)} \, ,
\end{equation}
for $\mathcal{R}>\mathcal{R}_{\text{b}}$ and $\Delta t_{\text{esc}}\left(\mathcal{R}\right) = t_{\text{life}}$ otherwise. Choosing $\mathcal{R}_{\text{b}} = \mathcal{O}\left(10\right) \, \si{\tera\volt}$ is motivated by the maximum rigidity that can be confined in SNR environments according to ~\citet{1983LagageCesarskyA&A}. Setting $\mathcal{R}_{\text{b}} \rightarrow \infty$ lets all protons escape at the same time. We will refer to this scenario, which is usually considered as the standard injection in stochastic modeling~\citep{Mertsch2011,Genolini2017,EvoliAmatoBlasi2021_Stochastic}, as burst-like injection.
In our model, individual sources are thus uniquely specified by their positions, ages, and values for $\mathcal{R}_{\text{b}}$.

It has been shown (see, e.g., \citet{2022RecchiaGalliA&A,2016NavaGabiciMNRAS,2019NavaRecchiaMNRAS,2022JacobsMertschJCAP}) that the escape of cosmic rays from the shock and residence in the source region is also influenced by the resonant streaming instability. However, the residence time due to this is falling quickly as a function of rigidity and is always smaller than the $t_{\text{esc}}$ considered here as long as $\mathcal{R}_{\text{b}} \gtrsim 100 \, \text{GV}$.
We take the simulations in \citep{2009CaprioliBlasiMNRAS1} and the considerations of \citet{1983LagageCesarskyA&A} as a motivation to choose $\mathcal{R}_{\text{b}}$ close to \SI{10}{\tera\volt}, which means that the overall time-dependence is hardly affected by this.
As the effect of individual sources on the cosmic-ray spectrum at low rigidities is small, we omit modifications of the time-dependent escape for $\mathcal{R} < \mathcal{R}_{\text{b}}$ that might occur in the late phase of the sources' lifetime and model them as burst-like injection at the time $t_{\text{life}}$.

The source injection term on the right-hand side of Equation~\eqref{eq:transport_equation} for a single source at $\left(t_i, \mathbf{x}_i\right)$ is given by
\begin{equation}
    Q_i\left(\mathcal{R}, t, \mathbf{x}\right) = Q_{\mathcal{R}} \left(\mathcal{R}\right) \delta\left(\mathbf{x} - \mathbf{x}_i\right) \delta\left(t - t_{\text{esc},i}\left(\mathcal{R}\right)\right) \, , 
\end{equation}
where $Q_{\mathcal{R}}(\mathcal{R})$ is the time-integrated source spectrum as described in Equation~\eqref{eq:time_inegrated_source_spectrum} and is the same for all sources. Motivated by diffusive shock acceleration~\citep{2001RPPh...64..429M}, it has a power-law dependence on rigidity given by $Q_{\mathcal{R}}(\mathcal{R}) \propto \mathcal{R}^{-2.2}$. Note that the deviation from the canonical $\mathcal{R}^{-2}$ spectrum can be explained as due to the relative speed of the scattering centers~\citep{Bell:1978ztp,Caprioli:2020spz} or shock obliquity~\citep{Bell:2011cs}.
As the source spectrum normalization only enters as an overall factor and we will only be interested in relative deviations of stochastic proton spectra from the ensemble mean, we do not need to fix it in this study. 

We solve the transport equation \eqref{eq:transport_equation} under the assumption of two free-escape boundary conditions, i.e., $\psi\left(z = \pm H \right) = 0$, using the method of mirror charges. We do not consider a radial boundary condition as we assume the observer to be sufficiently far away from the edge of the Galaxy such that the escape is dominantly in the $z$-direction.
The Green's function of Equation~\eqref{eq:transport_equation}, i.e., the solution for a single source, is given by 
\begin{equation}
    G\left(t, \mathbf{x}; t_i, \mathbf{x}_i\right) =  \frac{Q_{\mathcal{R}}\left(\mathcal{R}\right)}{\left(2 \pi \sigma^2\right)^{\frac{3}{2}}} e^{-\frac{\left(\mathbf{x}_i-\mathbf{x}\right)^2}{2 \sigma^2}} \vartheta\left(z, \sigma^2, H\right) \, ,
\end{equation}
where $\sigma^2\left(\mathcal{R}, t; t_i \right) = 2 \, \kappa\left(\mathcal{R}\right) \left(t - t_{\text{esc},i}(\mathcal{R})\right)$. The Green's function contains the power-law source spectrum, a Gaussian, and a correction function $\vartheta$, which accounts for the free-escape boundary condition. This function is an infinite sum related to the Jacobi theta function~\citep{Mertsch2011}. It has values in $\left[0,1\right]$ and gets much smaller than $1$ if $z \approx \pm H$ or if the diffusion length ($\approx \sigma$) is comparable with $H$.
We exclude source injections that lie outside the observer's past light-cone to remedy potential non-causal solutions of the diffusion equation~\citep{Genolini2017}.
Further cuts of young and close sources have been suggested (see, e.g., \citet{2012BlasiAmatoJCAP}). However, it is not at all clear how such a cut parameter should be chosen. While configurations with high fluxes from young or close sources might not be realized in our position in the Galaxy, they are inherent predictions of the model of individual cosmic-ray sources. Only if we consider the flux from all physically realizable configurations drawn from the source distribution do we find the unbiased model prediction for detecting features. For a discussion of using catalogs in stochastic modeling, see, e.g., \citet{2018MertschJCAP}.

Predictions for the local spectrum of GCR protons can be obtained by summing the Green's functions of tens of millions of sources, whose positions and ages are sampled from the respective distributions described above.
The embarrassingly parallel problem of calculating the contributions of all individual sources can be efficiently solved on graphical processing units (GPUs). 
To this end, we have used the \textsc{python} package \textsc{jax}~\citep{jax2018github}. 
Pronounced jump-like features originating in the time-dependent escape can be seen in sizable parts of the realizations. Locally measured GCR fluxes should thus reveal information about the escape of cosmic rays from sources.

\begin{table}[t]
\begin{tabular}{l l}
\hline\hline 
Parameter & Value \\
\hline
Spectral index of $\kappa\left(\mathcal{R}\right)$, $\delta$		& 0.6                    \\
Normalization of $\kappa\left(\mathcal{R}\right)$, $\kappa_0$	        &  \SI{6.8e27} {\centi\metre\squared\per\second} \\
Reference rigidity of $\kappa\left(\mathcal{R}\right)$, $\mathcal{R}_0$ &   \SI{1}{\giga\volt}\\
Galactic center distance, $R_{\odot}$			& \SI{8.3}{\kilo\parsec} \\
Galactic radius, $R$			& \SI{15}{\kilo\parsec}  \\
Halo height, $H$				& \SI{4}{\kilo\parsec}   \\
Sedov time, $t_{\text{Sed}}$		& \SI{1}{\kilo\year}   \\
Lifetime of sources, $t_{\text{life}}$	& \SI{100}{\kilo\year}   \\
Maximum injected $\mathcal{R}$, $\mathcal{R}_{\text{knee}}$           & \SI{3}{\peta\volt}\\
Source rate, $\nu$                 & \SI{0.03}{\per\year}\\
\hline\hline
\end{tabular}
\caption{Fiducial values for the GCR proton spectrum}
\label{tab:parameters}
\end{table}

%----------------------------------------------------------
%----------------------------------------------------------
%----------------------------------------------------------
\section{Results}

\begin{figure*}[!tbh]
\includegraphics[width=\linewidth]{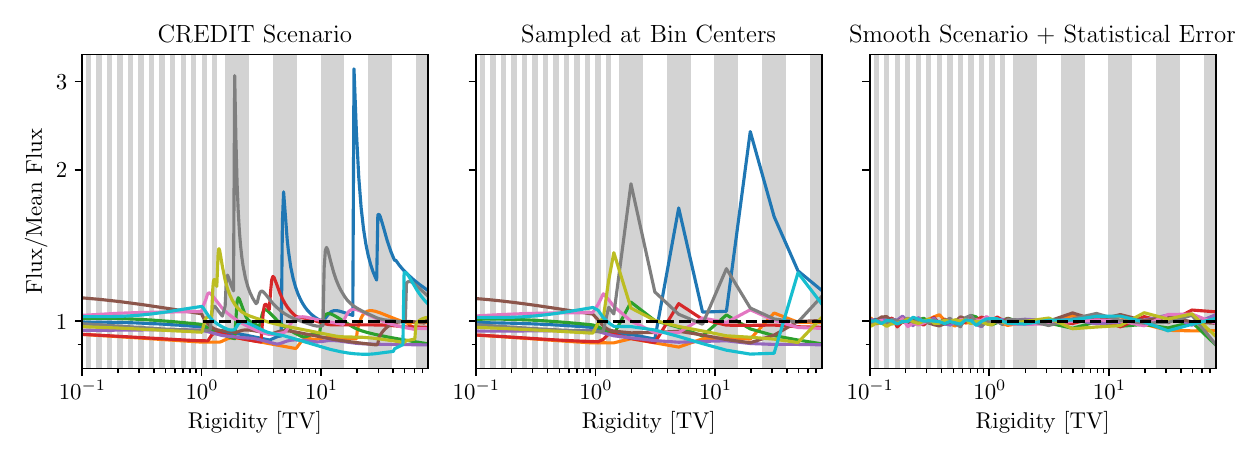}
\caption{GCR proton fluxes (normalized to the ensemble mean) predicted for the CREDIT scenario with $\mathcal{R}_{\text{b}} \in \left[1,25\right]\,\si{\tera \volt}$. The alternating white and gray vertical lines indicate the rigidity bins. \textbf{Left panel:} the colored lines show 10 random realizations, adopting a resolution in rigidity much higher than currently realized by experiments. \textbf{Middle panel:} same random realization but sampled at the centers of the rigidity bins employed by AMS-02 and DAMPE. \textbf{Right panel:} fluxes in a smooth scenario plus uncorrelated errors due to finite event numbers and systematic errors as published by AMS-02 and DAMPE.}
\label{Fig: panel realisations}
\end{figure*}

In the left panel of Figure~\ref{Fig: panel realisations}, we show a number of example spectra for a CREDIT scenario where we adopted the fiducial parameters presented in Table~\ref{tab:parameters}.
We present the spectra for ten random realizations divided by the ensemble mean for $\mathcal{R}_{\text{b}} \in \left[1,25\right]\,\si{\tera \volt}$. The parameter $\mathcal{R}_{\text{b}}$ is chosen for each source individually from a distribution $p\left(\mathcal{R}_{\text{b}}\right) \propto 1/\mathcal{R}_{\text{b}}$.
One can see fluctuations around the average spectrum for all rigidities, but their character changes markedly at around a rigidity of \SI{1}{\tera\volt}: 
For $\mathcal{R} \lesssim \SI{1}{\tera\volt}$, the spectral features are very smooth and typically only extend a few percent above or below the average value. 
For $\mathcal{R} \gtrsim \SI{1}{\tera\volt}$ instead, the spectral features are much more peaked, and the maxima of individual features can be as large as a few times the average spectrum in some realizations. Note that there are no sharp spectral features in burst-like injection; hence, these features can be used to discriminate between the CREDIT and the burst-like scenario.

Of course, given the limited event statistics, experimental bins are typically wider than these spectral features. 
In the middle panel of Figure~\ref{Fig: panel realisations}, we, therefore, present the same example spectra as in the left panel but sampled only at the central rigidities of bins employed by experimental collaborations. 
Specifically, for rigidities below and above $\approx \SI{2}{\tera\volt}$, we have adopted the published rigidity bins of the AMS-02 and DAMPE collaborations, respectively. 
While the features in the DAMPE range are not as high and as nicely resolved as before, the most dramatic spectral features are still easy to make out by eye. 
We checked that integrating a finely resolved spectrum over rigidity bins to better model the binning process of the experiments gives similar statistics of the fluxes. 

Statistical errors also arise due to the finite number of registered events. 
For a large enough number of events per bin, the distribution of the measured fluxes will be Gaussian, and the measured fluxes in different bins can be considered as independent. 
We use reported statistical errors of the calorimetric cosmic-ray experiments AMS-02~\citep{2021AguilarAliCavasonzaPhR} and DAMPE~\citep{2019AnAsfandiyarovSciA} as standard deviations in the individual rigidity bins. We found that the relation $\sigma_{\text{stat}}[\%] = 0.42 \left(\mathcal{R}[\si{\tera\volt}]\right)^{0.6}$ approximates the statistical errors of both experiments well.
We add the effect of the uncorrelated part of the systematic error by adding in quadrature an uncorrelated, relative systematic error of $1\%$ to the purely statistical errors~\citep{Cavasonza:2016qem,AMS:2015tnn}.
In the right panel of Figure~\ref{Fig: panel realisations}, we show the spectrum from the smooth model divided by the average, a flat line, plus ten random draws from a distribution of uncorrelated errors only. 
Generally, the statistical fluctuations are much smaller than the expected features under the CREDIT scenario, rendering their detection possible. 

\begin{figure}[!tb]
\includegraphics[width=\linewidth]{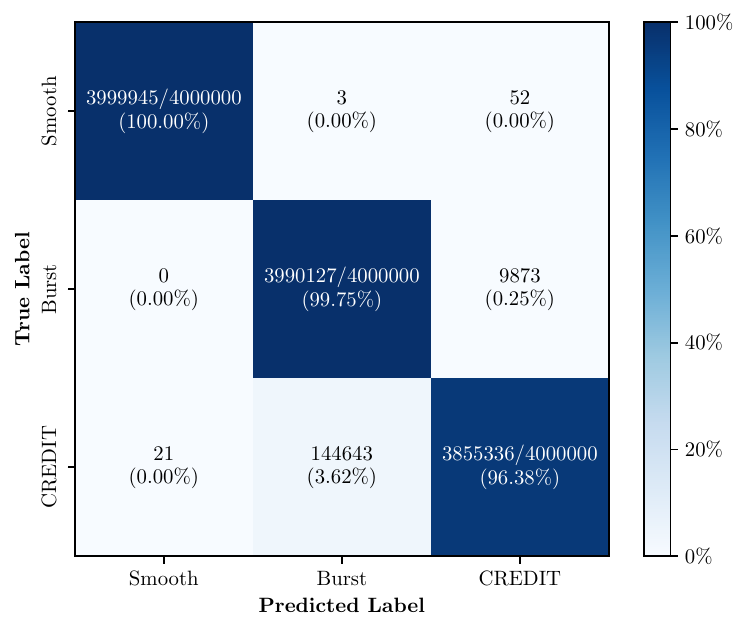}
\caption{Confusion matrix (normalized to rows for the percentages) where the CREDIT realizations are simulated with $20$ distinct, but for each realization fixed $\mathcal{R}_{\text{b}} \in \left[1,25\right]\,\si{\tera\volt}$.}
\label{fig2:confusion_matrix_separate}
\end{figure}

\begin{figure}[!tb]
\includegraphics[width=\linewidth]{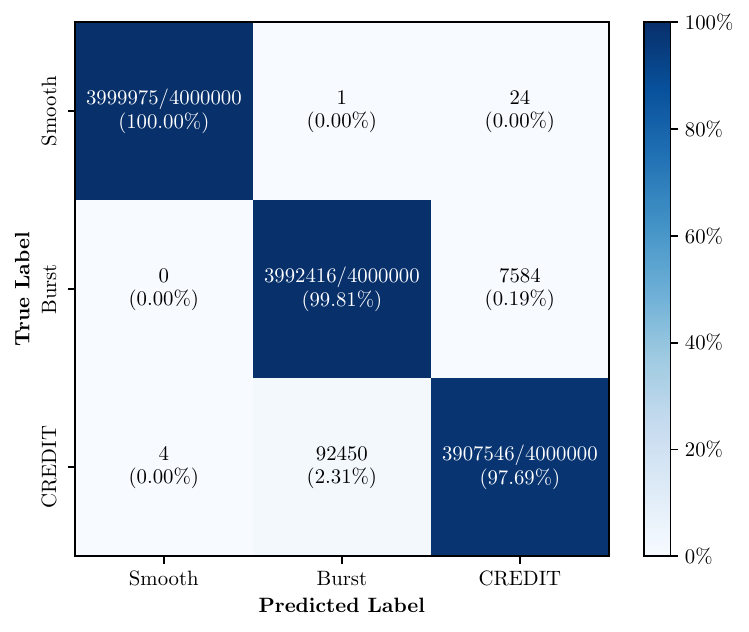}
\caption{Same as Figure~\ref{fig2:confusion_matrix_separate}, but $\mathcal{R}_{\text{b}}$ is chosen randomly for each individual source from a distribution $p\left(\mathcal{R}_{\text{b}}\right) \propto 1/\mathcal{R}_{\text{b}}$ for $\mathcal{R}_{\text{b}} \in \left[1,25\right]\,\si{\tera\volt}$ for the CREDIT model.}
\label{fig3:confusion_matrix_mixed}
\end{figure}

However, it is conceivable that statistical fluctuations can produce a flux similar to a realization with less dramatic features. 
We trained a decision tree (DT) to discriminate between the null hypothesis of a smooth source distribution (plus uncorrelated errors) and the alternative hypotheses of a CREDIT ($\mathcal{R}_{\text{b}} \in \left[1,25\right]\,\si{\tera\volt}$)\footnote{We chose values close to the maximum confinement rigidity of \SI{10}{\tera\volt} suggested by \citet{1983LagageCesarskyA&A}. The upper value of \SI{25}{\tera\volt} is motivated by the requirement of having at least three bins at higher rigidities for spectral features to appear.} or a burst-like ($\mathcal{R}_{\text{b}}\rightarrow\infty$) scenario.
The DT aims to split the data, the flux value in each rigidity bin above \SI{10}{\giga\volt}, scattered in 58-dimensional space, into hypercuboids that contain realizations of the same label~\citep{bishop2006pattern}. This is done by continuously splitting one hypercuboid (and the data set) into two smaller hypercuboids.
This can be thought of as an if-statement, where if the value of a realization at the bin is smaller than a fixed value, it is in one hypercube and otherwise in the other.
Making a prediction for a new data point is just traveling through a number of if-statements and obtaining the label at the end. We used the \textsc{sklearn} package~\citep{scikit-learn} to train and validate our DT. 
No tuning of hyperparameters was required to improve the results.
For the training, we used realizations of the three classes smooth, burst, and CREDIT and trained the DT with their respective fluxes at rigidities in the range from around \SI{10}{\giga\volt} to \SI{100}{\tera\volt}. For the CREDIT realizations, we tried two ways to vary the parameter $\mathcal{R}_{\text{b}}$. In one case, we fixed $\mathcal{R}_{\text{b}}$ for each individual realization and trained the DT on a mixture of $20$ different $\mathcal{R}_{\text{b}} \in \left[1,25\right]\,\si{\tera\volt}$, and in the other, we chose $\mathcal{R}_{\text{b}}$ randomly from the interval $\left[1,25\right]\,\si{\tera\volt}$ for each source, as described at the beginning of this section.

We estimate the significance of our hypothesis test and its power through the so-called confusion matrix. 
The confusion matrix shows how often an element of the validation set with a true label is classified with a predicted label. 
A diagonal matrix corresponds to a perfect classification. Using 10-fold cross-validation on our $4 \times 10^6$ realizations for each scenario, we find almost diagonal confusion matrices (Figures~\ref{fig2:confusion_matrix_separate} and \ref{fig3:confusion_matrix_mixed}) which shows the strength of DTs to reliably distinguish the three models. The two ways of varying $\mathcal{R}_{\text{b}}$ have little influence on the result.

The fiducial parameters given in Table~\ref{tab:parameters}  do not have to be realized exactly like this. The parameter influence should be studied in more detail in future work. For now, we just checked that halving the source rate $\nu$ or the sources' lifetime $t_{\text{life}}$ does not change the confusion matrix structure significantly.
We also considered a relaxation of the peaked injection of each rigidity at a single point in time and found that smoothing the injection by a Gaussian kernel with a conservative width of \SI{1}{\kilo\year} does not alter the results for $\mathcal{R}_{\text{b}} \gtrsim \SI{1}{\tera\volt}$.

%----------------------------------------------------------
%----------------------------------------------------------
%----------------------------------------------------------
\section{Summary and conclusion}

We considered a stochastic source model for simulating the GCR proton spectrum.  At rigidities above some teravolts, the spectrum is determined by relatively few sources.
We discussed that rigidity-dependent escape times of protons from their sources (CREDIT scenario) enhance the observability of single source contributions in the spectrum. Such an escape model, motivated by the same physical mechanisms that are usually invoked to explain particle acceleration in Galactic SNRs beyond $\approx \SI{10}{\tera\volt}$, predicts a high likelihood for the emergence of peaked features in the proton spectrum. Those peaks are characteristic for CREDIT models (see Figure~\ref{Fig: panel realisations}).
Here, we used a DT to discriminate spectra produced by a CREDIT or burst-like source injection from a power-law spectrum and a smooth source distribution. The classifier works extremely well, reliably deciding whether a spectrum is compatible with a smooth source injection or a CREDIT/burst-like model. 

We suggest applying our classifier to experimental data in the future. 
If the data get classified as CREDIT-like, one can determine the time of the supernova event $t_i$ for an observed spectral feature if other model parameters are assumed, i.e., diffusion coefficient $\kappa(\mathcal{R})$, break rigidity $\mathcal{R}_{\text{b}}$, Sedov time $t_{\text{Sed}}$, and lifetime of sources $t_{\text{life}}$. 
Alternatively, for a candidate supernova time $t_i$, one can constrain a combination of $\kappa(\mathcal{R})$, $\mathcal{R}_{\text{b}}$, $t_{\text{Sed}}$, and $t_{\text{life}}$ that fits a detected spectral feature. 
Instead, if the data get classified as burst-like, this corresponds to a lower bound on $\mathcal{R}_{\text{b}}$, which informs models of acceleration. 
Finally, if the classifier finds a smooth model, this should trigger a reckoning with the SNR paradigm for GCRs as the non-detection of source signatures would be very unlikely within this paradigm. 

For this, however, the cosmic-ray model parameters have to be chosen carefully such that no correlated spectral features attributed to the transport model emerge in the normalized spectra (like Figure~\ref{Fig: panel realisations}). This can be done using extensive fits like, e.g.,~\citet{Schwefer2023_CRINGE}. Deviations from the fiducial values given in Table~\ref{tab:parameters} will change the spectral features, enhancing them whenever the effective number of contributing sources decreases (e.g., smaller halo height or enhanced diffusion).
A detailed investigation of the parameter influences requires a more extended study.

Producing extensive Monte Carlo simulation samples using state-of-the-art programming paradigms opens up the possibility to train classifiers to reliably distinguish features of different models in data. We expect this to be a powerful tool to constrain GCR models. The possibilities of this will be explored in future work but are already demonstrated by the classification power to discriminate CREDIT against burst-like models.

As the number of contributing sources falls with rigidity, extensions of high-precision measurements beyond~$\approx$~\SI{100}{\tera\volt} can enhance the predictive power of the classifiers. One proposed space-based direct detection experiment is AMS-100~\citep{2019NIMPA.94462561S} which prospectively will extend high-precision proton measurements beyond the cosmic-ray knee. The leading high-precision direct detection experiment DAMPE is currently limited by statistics and suffers from increasingly limited power to discriminate proton and helium at rigidities beyond~$\approx$~\SI{100}{\tera\volt}~\citep{2024AlemannoAltomarePhRvD}. The inclusion of helium spectra in our model and the use of the p+He spectra provided by DAMPE can also increase the predictive power of our model.
In addition, an unbinned analysis would further improve the sensitivity to the spectral features. 
We postpone such a study to future work.

\section*{Acknowledgements}
This project was funded by the Deutsche Forschungsgemeinschaft (DFG, German Research Foundation) -- project number 490751943.

\bibliography{CREDIT.bib}
\end{document}